\documentclass[10pt,aps,prb,twocolumn,amsmath,amssymb,superscriptaddress]{revtex4-1}
\usepackage{amsmath}
\usepackage{graphicx}
\usepackage{xcolor}
\usepackage{multirow}
\usepackage[caption = false]{subfig}
\usepackage[normalem]{ulem}
\usepackage{epstopdf}

\usepackage{algcompatible}
\usepackage[floatrow]{trivfloat}
\trivfloat{algorithm}

\def\beq{\begin{eqnarray}}
\def\eeq{\end{eqnarray}}
\def\beqq{\begin{eqnarray*} \color{blue} }
\def\eeqq{\end{eqnarray*}}

\def\V{\mathcal{V}}
\def\P{\mathcal{P}}

\begin{document}

\title{Orbital optimization in selected configuration interaction methods}

\author{Yuan Yao}
\author{C. J. Umrigar}
\affiliation{
  Laboratory of Atomic and Solid State Physics,\\
  Cornell University, Ithaca, NY 14853.}

\begin{abstract}
We study several approaches to orbital optimization in selected configuration interaction plus perturbation theory (SCI+PT) methods, and test them on the ground and excited states of three molecules using the semistochastic
heatbath configuration interaction (SHCI) method. We discuss the ways in which the orbital optimization problem in SCI resembles and differs from that in complete active space self-consistent field (CASSCF).
Starting from natural orbitals, these approaches divide into three classes of optimization methods according to how they treat coupling between configuration interaction (CI) coefficients and orbital parameters, namely uncoupled, fully coupled, and quasi-fully coupled
methods. We demonstrate that taking the coupling into account is crucial for fast convergence and recommend two quasi-fully coupled
methods for such applications: accelerated diagonal Newton and Broyden-Fletcher-Goldfarb-Shanno (BFGS).

\end{abstract}

\maketitle

\section{Introduction}

Selected configuration interaction plus perturbation theory (SCI+PT) methods are an important class of
electronic structure theory methods applicable to both weakly and strongly correlated
systems~\cite{HurMalRan-JCP-73,BuePey-TCA-74,EvaDauMal-CP-83,GinSceCaf-CJC-13,Eva-JCP-14,SceAppGinCaf-JCoC-16,GarSceLooCaf-JCP-17,LooSceBloGarCafJac-JCP-18,HaiTubLevWhaHea-JCTC-19,LooLipPasSceJac-JCTC-20}.
The original CIPSI (Configuration Interaction using a Perturbative Selection made Iteratively)
method dates back almost 50 years~\cite{HurMalRan-JCP-73},
but recent improvements to this method and the development of new flavors have made SCI the
method of choice for some chemical systems.
These methods consist of two stages: In the first stage a variational wave function is constructed iteratively, starting from
a determinant that is expected to have a significant amplitude in the final wave function, e.g., the Hartree-Fock (HF)
determinant.
In the second stage, second-order perturbation theory is used to improve upon the variational energy.
Compared to complete active space self-consistent field (CASSCF) methods which are also frequently applied to treat strong correlation,
SCI methods do not necessarily require the prior specification of an active space and instead can
select the most important determinants from the entire Hilbert space.

Although orbital optimization is an integral part of the CASSCF procedure and much effort has been devoted to its study, it remains inadequately explored in the SCI literature.
Orbital optimization in the two methods share a fair amount of similarity: both methods employ a
configuration interaction (CI) expansion for the wave function and both methods optimize both
CI coefficients and orbital coefficients.
For this reason, the optimization strategies we present here will borrow heavily from  progress made in the CASSCF studies.~\cite{WerMey-JCP-80,WerKno-JCP-85,SunYanCha-CPL-17,KreWerKno-JCP-19,KreWerKno-JCP-20}


However, major differences also exist between the two methods. CASSCF categorizes the molecular orbitals into inactive, active, and virtual orbitals based on occupation and
includes all determinants formed by exciting from and to orbitals in the active space.
For this reason, internal rotations of orbitals within each of the three spaces are redundant, i.e.,
they do not change the energy.
In contrast, categorization of orbitals is not necessary in SCI methods, and in this paper we
will not categorize them, i.e., all orbitals may or may not be occupied in some determinant.
In other words, all orbitals are active and all orbital rotations are relevant.
This makes the orbital optimization problem harder, not only because there are more parameters
but also because there is stronger coupling between the orbital and the CI parameters.

Most SCI calculations to date have been performed using the following types of orbitals: canonical HF orbitals, localized orbitals,
CASSCF orbitals, and natural orbitals which are eigenstates of the one-body reduced density matrix (1-RDM)---the 1-RDM can be obtained from either an initial SCI wave function constructed with HF orbitals or a wave function from
some other theory, such as coupled cluster with single and double excitations (CCSD).
These latter choices typically produce wave functions that have faster convergence with respect to the number of determinants.
However, as we demonstrate in this paper, much can be gained from further optimizing these orbitals to minimize the SCI variational energy itself.
Although we only consider extended orbitals in this study, the same procedure can be applied to localized orbitals as well.


We demonstrate the effectiveness of the optimization procedures we propose using the semistochastic heatbath
configuration interaction (SHCI) method, but the same procedures can be applied within other SCI methods as well.
There has been prior work~\cite{SmiMusHolSha-JCTC-17,LevHaiTubLehWhaHea-JCTC-20} on optimizing orbitals in SCI,
but in these papers only excitations in a limited active space were allowed whereas we allow all excitations.
We note that optimizations
of both variational energy and total energy (variational plus perturbative energies) were
considered in Ref.~\onlinecite{SmiMusHolSha-JCTC-17}.  Here, we target just the variational energy but
study the effect of the optimization on both variational and total energies.

As has been observed in CASSCF settings, second-order optimization in the orbital subspace alone does not
result in rapid convergence due to substantial coupling between the CI and orbital subspaces.
To achieve rapid convergence one must take this coupling into account.
Therefore, the optimization strategies naturally fall into three classes: uncoupled optimization which only performs second-order optimization in the orbital subspace, fully coupled optimization which explicitly couples the two subspaces at prohibitive computational cost, and what we call quasi-fully coupled schemes which represent various compromises between the previous two.

The rest of this paper is organized as follows. In Section \ref{SHCI_review} we review the basic procedure of SHCI and set the notation. In Section \ref{optimization_strategies} we present the various optimization strategies. We demonstrate their performance on some representative systems and illustrate their pros and cons. The details of the  efficient evaluation of the various quantities used in optimization are left for Section \ref{evaluation_pieces}. In Section \ref{conclusions} we conclude this study.

\section{Review of SHCI}
\label{SHCI_review}

The second-quantized nonrelativistic electronic Hamiltonian in the Born-Oppenheimer approximation is
\begin{align}
\label{qc_hamiltonian}
\hat{H} &=\hat{h} + \hat{g} + h_{\rm nuc} \nonumber \\
&=\sum_{pq}h_{pq}\hat{E}_{pq}+\frac{1}{2}\sum_{pqrs} g_{pqrs}\hat{e}_{pqrs} + h_{\rm nuc},  
\end{align}
where the one- and two-electron excitation operators are
\begin{align}
\label{1b_excitation}
\hat{E}_{pq} = \sum_{\sigma}{a_{p\sigma}^{\dagger} a_{q\sigma}}
\end{align}
and
\begin{align}
\label{2b_excitation}
\hat{e}_{pqrs} = \sum_{\sigma\tau}{a_{p\sigma}^{\dagger} a_{r\tau}^{\dagger} a_{s\tau}a_{q\sigma}},
\end{align}
respectively. Here, $p,q,r,s \in \{1,2,\ldots,N_{\rm orb}\}$ are the molecular orbital indices, and $\sigma,\tau \in \{\uparrow,\downarrow\}$ are the spin indices.

The corresponding one- and two-electron integrals are defined in terms of the spatial molecular orbitals $\phi({\bf r})$ as:
\begin{align}
h_{pq} &= \int \phi_p^*({\bf r}) \left(-\frac{1}{2} \nabla^2 -\sum_{I} \frac{Z_I}{\lvert {\bf r} - {\bf r}_I\rvert} \right) \phi_q({\bf r})
d{\bf r} \label{1b_integrals}\\
g_{pqrs} &= \iint \frac{\phi^*_p({\bf r}_1)\phi^*_r({\bf r}_2)\phi_q({\bf r}_1)\phi_s({\bf r}_2) }{\lvert{\bf r}_1 - {\bf r}_2\rvert} d{\bf
r}_1 d{\bf r}_2 \label{2b_integrals},
\end{align}
where $Z_I$ and ${\bf r}_I$ are the nuclear charges and coordinates. The nuclear repulsion energy is
\begin{align}
h_{\rm nuc} = \sum_{I<J} \frac{Z_I Z_J}{\lvert {\bf r}_I -{\bf r}_J \rvert}.
\end{align}

SHCI finds near exact solutions for the ground and excited states of the electronic Hamiltonian in Eq. (\ref{qc_hamiltonian}) in a two-stage
procedure: the variational stage and the perturbative stage.
In the following, we use $\V$ for the set of variational determinants, and $\P$ for the set of perturbative
determinants, that is, the set of determinants that are connected to the variational determinants by at least one
non-zero Hamiltonian matrix element but are not present in $\V$.
\subsection{Variational stage}
\label{Var}
Like all SCI methods, SHCI starts from an initial determinant such as the HF determinant
and generates the variational wave function, $\Psi_V$, through an iterative procedure.

At each iteration, $\Psi_V$ is written as a linear combination of the determinants
in the space $\V$
\begin{align}
\label{variational_wf}
\left|\Psi_{V} \right\rangle= \sum_{D_i \in \V} c_{i} \left|D_{i}\right\rangle
\end{align}
and new determinants, ${D_a}$, from the space $\P$ that satisfy the criterion
\beq
\exists\; D_i \in \V , \mathrm{\ such\ that\ } \left|H_{a i} c_{i}\right| \ge \epsilon_{1}
\label{HCI_criterion}
\eeq
are added to the $\V$ space, where
$H_{ai}$ is the Hamiltonian matrix element between determinants $D_a$ and $D_i$, and
$\epsilon_1$ is a user-defined parameter that controls the accuracy of the variational
stage~\footnote{Since the absolute values of $c_i$ for the most important determinants tends to go down as more
determinants are
included in the wave function, a somewhat better selection of determinants is obtained by using a larger value of
$\epsilon_1$ in the initial iterations.}.
(When $\epsilon_1=0$, the method becomes equivalent to FCI.)
After adding the new determinants to $\V$, the Hamiltonian matrix is constructed and diagonalized using the diagonally
preconditioned Davidson method~\cite{Dav-CPC-89} to obtain an improved estimate of the lowest eigenvalue, $E_{V}$,
and eigenvector, $\Psi_V$.
This process is repeated until the change in the variational energy $E_V$ falls below a certain threshold.

Other SCI methods use different criteria, based on
either the first-order perturbative coefficient of the wave function,
\beq
\left|c_a^{(1)}\right|=\left|\frac{\sum_i H_{ai}c_i}{E_V-E_a}\right| > \epsilon_1,
\label{eq:cipsi_wf}
\eeq
or the second-order perturbative correction to the energy,
\beq
-\Delta E^{(2)}=-\frac{\left(\sum_i H_{ai}c_i\right)^2}{E_V-E_a} > \epsilon_1,
\label{eq:cipsi_energy}
\eeq
where $E_a = H_{aa}$.
The reason SHCI uses instead the selection criterion in Eq.~(\ref{HCI_criterion}) is that it can be implemented
very efficiently without checking the vast majority of the determinants that do not meet the criterion, by taking
advantage
of the fact that
most of the Hamiltonian matrix elements correspond to double excitations and that their values do not depend
on the determinants themselves but only on the four orbitals whose occupancies change during the double excitation.
Therefore, at the beginning of an SHCI calculation, for each pair of spin-orbitals, the absolute values of the Hamiltonian
matrix elements obtained by doubly exciting from that pair of orbitals is computed and stored
in decreasing order by magnitude, along with the corresponding pairs of orbitals the electrons would excite to.
Then the double excitations that meet the criterion in Eq.~(\ref{HCI_criterion}) can be generated by
looping over all pairs of occupied orbitals in the reference determinant and
traversing the array of sorted double-excitation matrix elements for each pair.
As soon as the cutoff is reached, the loop for that pair of occupied orbitals is exited.

Although the criterion in Eq.~(\ref{HCI_criterion}) does not include information from the diagonal elements,
the selected determinants are not much different from those selected by either of the criteria
in Eqs.~(\ref{eq:cipsi_wf}) and (\ref{eq:cipsi_energy}) because
the terms in the numerators of Eqs.~(\ref{eq:cipsi_wf}) and (\ref{eq:cipsi_energy})
span many orders of magnitude, so the sums are highly correlated with the largest-magnitude term in the sums in
Eqs.~(\ref{eq:cipsi_wf}) or (\ref{eq:cipsi_energy}), and because the denominator is never small
after several determinants have been included in $\V$.
It was demonstrated in Ref.~\onlinecite{HolTubUmr-JCTC-16} that the selected determinants give only slightly
inferior convergence
to those selected using the criterion in Eq.~(\ref{eq:cipsi_wf}).  This is greatly outweighed by the improved
selection speed.
Moreover, one could use the criterion in Eq.~(\ref{HCI_criterion}) with a smaller value of $\epsilon_1$ as a
preselection criterion, and then select determinants
using the criterion in Eq.~(\ref{eq:cipsi_energy}) or something close to it, thereby having the benefit of both a
fast selection method and a
close to optimal choice of determinants.
We use a similar but somewhat more complicated criterion for the selection of the determinants
connected to those in $\V$ by a single excitation as well, but this improvement is of lesser importance
because there are many fewer singly excited connections than doubly excited ones.
With these improvements the time required for selecting determinants is negligible, and
the most time consuming step by far in the variational stage is the construction of the sparse
Hamiltonian matrix.  Details for doing this efficiently are given in Ref.~\onlinecite{LiOttHolShaUmr-JCP-18}.

When computing excited states in addition to the ground state\cite{HolUmrSha-JCP-17}, all states are expanded in the same set of variational determinants
\begin{align}
\left|\Psi_{V}^{(s)} \right\rangle= \sum_{D_i \in \V} c_{i}^{(s)} \left|D_{i}\right\rangle,
\end{align}
where $s$ indexes the state. At each iteration, we add to $\V$ the union of the new determinants that are important for each of the states. Thus, the determinants selection criterion in Eq. (\ref{HCI_criterion}) becomes
\beq
\left|H_{a i} \right| \left(\max_{s} \left|c_{i}^{(s)}\right| \right)\ge \epsilon_{1}.
\eeq
The Hamiltonian matrix constructed this way is then diagonalized using the Davidson method for as many of the lowest eigenvalues and eigenvectors as are desired.

\subsection{Perturbative stage}
\label{PT}
In common with most other SCI+PT methods, the perturbative correction is
computed using Epstein-Nesbet perturbation theory~\cite{Eps-PR-26,Nes-PRS-55}.
The variational wave function is used to define the zeroth-order Hamiltonian $\hat{H}^{(0)}$ and the perturbation $\hat{H}^{(1)}$:
\begin{align}
\hat{H}^{(0)} &= \sum_{D_i,D_j \in \V} H_{ij} |D_i\rangle\langle D_j| + \sum_{D_a \notin \V } H_{aa} |D_a\rangle\langle
D_a|. \nonumber\\
\hat{H}^{(1)} &= \hat{H} - \hat{H}^{(0)} . \label{eq:part}
\end{align}
The first-order energy correction is zero, and the second-order energy correction $\Delta E^{(2)}$ is
\begin{align}
\Delta E^{(2)} = \langle\Psi_V|\hat{H}^{(1)}|\Psi^{(1)}\rangle
\;=\; \sum_{D_a \in \P} \frac{\left(\sum_{D_i \in \V} H_{ai} c_i\right)^2}{E_V - E_a},
\label{eq:PTa}
\end{align}
where $\Psi^{(1)}$ is the first-order wave-function correction.
The SHCI total energy is then
\beq
\label{eq:E_tot}
E^{\rm SHCI} &=& E_V + \Delta E^{(2)} \;=\; \langle {\Psi_V}|\hat{H}|{\Psi_V} \rangle + \Delta E^{(2)}.
\eeq

It is expensive to evaluate the expression in Eq.~(\ref{eq:PTa}) because the outer summation includes all determinants
in the space $\P$ and their number is
${\cal O}(N_\text{elec}^2 N_\text{unocc}^2 N_{\rm det})$, where $N_{\rm det}$ is the number of variational determinants in $\V$, $N_\text{elec}$ is the number of electrons,
and $N_\text{unocc}$ is
the number of unoccupied orbitals.
The straightforward
and time-efficient approach to computing the perturbative correction requires storing
the partial sum $\sum_{D_i \in \V} H_{ai} c_i$ for each unique $a$, while
looping over all the determinants $D_i \in \V$. This creates a severe memory bottleneck.
A widely used alternative approach does not require storing the unique $a$, but requires checking whether the determinant has
already been generated by checking its connection with variational determinants whose connections have already been included.
This entails some additional computational expense.

The SHCI algorithm instead uses two other strategies to reduce both the computational time and the storage requirement.
First, SHCI screens the sum~\cite{HolTubUmr-JCTC-16} using a second threshold, $\epsilon_2$ (where
$\epsilon_2<\epsilon_1$) as the criterion for selecting perturbative determinants $D_a \in \P$,
\begin{equation}
\Delta E^{(2)} \left(\epsilon_{2}\right) = \sum_{D_a \in \P} \frac{\left(\sum_{D_i \in \V}^{(\epsilon_{2})}H_{a i} c_{i}\right) ^{2}}{E_{V} - E_a}
\label{eq:PTb}
\end{equation}
where $\sum^{(\epsilon_{2})}$ indicates that only terms in the sum for which $\left|H_{a i} c_{i}\right| \ge
\epsilon_{2}$ are included.
Similar to the variational stage, we find the connected determinants efficiently with precomputed arrays of
double excitations sorted by the magnitude of their Hamiltonian matrix elements~\cite{HolTubUmr-JCTC-16}.
Note that the vast number of terms that do not meet this criterion are \emph{never evaluated}.
Even with this screening, the simultaneous storage of all terms indexed by $a$ in Eq.~(\ref{eq:PTb}) can exceed
computer memory
when $\epsilon_2$ is chosen small enough to obtain essentially the exact perturbation energy.
The second innovation in the calculation of the SHCI perturbative correction is to overcome this memory bottleneck
through semistochastic evaluation:
the most important contributions are evaluated deterministically and the rest are sampled stochastically.
Our original method used a two-step perturbative algorithm~\cite{ShaHolJeaAlaUmr-JCTC-17}, but our
later three-step perturbative algorithm~\cite{LiOttHolShaUmr-JCP-18} is even more efficient.
The three steps are:
\begin{enumerate}
  \item A deterministic step with cutoff $\epsilon_2^{\rm dtm} (<\epsilon_1)$, wherein
  all variational determinants are used, and
  all perturbative batches are summed over.
  \item A ``pseudo-stochastic" step with cutoff $\epsilon_2^{\rm psto} (< \epsilon_2^{\rm dtm})$, wherein
  all variational determinants are used, but the perturbative determinants are partitioned
  into statistically identical batches using a hash function.
  Typically, only a small fraction of these batches
  need to be summed over to achieve an error much smaller than the target error.
  \item A stochastic step with cutoff $\epsilon_2 (<\epsilon_2^{\rm psto}) $, wherein a few stochastic
  samples of variational determinants,
  each consisting of $N_d$ determinants, are sampled with probability $\vert c_i \vert/\sum_{D_i \in \V} \vert c_i \vert$,
  and only one of the perturbative batches is randomly selected per variational sample.
\end{enumerate}

We note that, subsequent to our first semistochastic paper~\cite{ShaHolJeaAlaUmr-JCTC-17}, a completely different, but also efficient,
semistochastic approach
has been presented in Ref.~\onlinecite{GarSceLooCaf-JCP-17}.

In a typical SHCI calculation, the variational energy and the corresponding perturbative correction are computed for several values of
$\epsilon_1$. To estimate the FCI energy, we perform a weighted quadratic fit of
$E^{\rm SHCI}$ to $-\Delta E^{(2)}$ to obtain $E^{\rm SHCI}$ at $-\Delta E^{(2)}=0$, using weights proportional to $(\Delta E^{(2)})^{-2}$. In
order to reduce the extrapolation error, one can either go to larger variational wave functions by decreasing $\epsilon_1$ or optimize the
orbitals to obtain better quality variational wave functions for the same number of determinants. The former approach will incur a larger
memory footprint and is limited in practice by the amount of computer memory available. In the next section, we discuss various orbital optimization methods.

\section{Optimization Strategies}
\label{optimization_strategies}

The optimization problem in SHCI can be formulated as minimizing the variational energy with respect to both CI coefficients $\bf c$ in Eq. (\ref{variational_wf})
and orbital parameters in $\bf X$:
\begin{align}
\label{energy2}
E({{\bf c}, {\bf X}}) = \frac{\langle \Psi_V | \exp(\hat{{\bf X}}) \hat{H} \exp(-\hat{{\bf X}}) | \Psi_V \rangle}
{\langle \Psi_V | \Psi_V\rangle}.
\end{align}
Here, $\hat{\bf X}$ is an antisymmetric one-electron operator
\begin{align}
\hat{\bf X} = \sum_{p>q} {\bf X}_{pq} \hat{E}_{pq}^{-},
\end{align}
and the antisymmetric singlet excitation operator is defined in terms of the one-body operator in Eq. (\ref{1b_excitation}) as
\begin{align}
\label{orbital_excitation}
\hat{E}_{pq}^{-} = \hat{E}_{pq} - \hat{E}_{qp}.
\end{align}
We use $\bf X$ to denote the $N_{\rm orb}\times N_{\rm orb}$ antisymmetric matrix so that $\exp(-\bf X)$ is an orthogonal matrix in orbital
space. (We only consider real rotations and real CI coefficients in this study, although the orbitals themselves are allowed to be real or complex.) The elements in the
triangular part of $\bf X$, say the upper triangle, are the set of unique orbital rotation parameters with respect to which we minimize the variational energy, and their number, $N_{\rm param}$, scales as $O(N_{\rm orb}^2)$. If
point group symmetry is used in the computation, then only elements of $\bf X$ corresponding to molecular orbitals of the same irreducible
representation  will be nonzero. Later, we will also use $\bf x$ to denote the $N_{\rm param} \times 1$ vectorized form of the $N_{\rm param}$
nonzero elements in the triangular part of  $\bf X$, say the upper triangular part. The number of nonredundant CI parameters is $N_{\rm det}-1$ since wave function normalization is implicitly enforced by the denominator of Eq. (\ref{energy2}).

A natural starting point for orbital optimization is natural orbitals. They are the eigenstates of the
1-RDM and represent orbitals with
definite occupation numbers (between 0 and 2) for
some wave function.
In the examples shown in this paper, we start from HF orbitals and then compute an SHCI wave function,
its 1-RDM, and natural orbitals as the first step of optimization.

We note that in some CASSCF methods
the orthogonal orbital rotation matrix is parameterized not as the exponential
of an antisymmetric matrix, $\exp(-{\bf X})$, but rather as the identity matrix plus a small correction matrix, $\bf 1+T$, with additional constraints on $\bf T$ to enforce orthogonality.~\cite{WerMey-JCP-80,WerKno-JCP-85,KreWerKno-JCP-19,KreWerKno-JCP-20}
The advantage of such a parameterization is that a second-order expansion of Eq. (\ref{energy2}) includes  
all orders of the orbital parameters in $\bf T$, thereby increasing the radius of convergence of second-order methods. 
We  have not found it necessary to resort to this parameterization, as the natural orbitals computed from the initial SHCI wave function typically constitute a good starting point and consequently the norm of $\bf X$ tends to be small.
Since natural orbitals tend to be much closer to the optimal set of orbitals than the starting HF orbitals,
we find very few, if any, negative eigenvalues in the orbital Hessian matrix.

In the rest of this section, we introduce various optimization strategies falling into three categories: uncoupled optimization where the coupling
between CI and orbital parameters is not taken into account, fully coupled optimization which explicitly calculates the entire Hessian matrix,
and quasi-fully coupled methods that implicitly take coupling into account at reduced computational cost.

We illustrate the performance of the various optimization strategies on three representative systems from recent benchmark calculations: H$_2$CO in the cc-pVTZ basis\cite{YaoGinLiTouUmr-JCP-20}, ScO in the
vdz basis using the Trail-Needs effective core potential\cite{WilYao_etal_UmrWag-PRX-20}, and Cr$_2$ in the cc-pVDZ-DK basis using the x2c Hamiltonian\cite{LiYaoHolOttShaUmr-PRR-20}. All these
molecules are in their ground state equilibrium  geometry. They represent molecules of increasing correlation strength. More details on these
systems are listed in Table~\ref{tab:molecule_info}.
These systems have fewer than 100 orbitals, but the largest systems we have optimized in an earlier work
have over 500 orbitals~\cite{YaoGinLiTouUmr-JCP-20}.
In fact, for the same system, larger basis sets show greater gains from orbital optimization.

\begin{table}
  \caption{Information on the three representative systems used in this section: system name, basis, point group used, and number of orbitals,
  electrons, and orbital parameters. All three systems are in their  ground state equilibrium geometry.
  }
  \begin{ruledtabular}
    \begin{tabular}{llrrrr}
      System & Basis & Pt Grp & $N_{\rm orb}$ & $N_{\rm elec}$ & $N_{\rm param}$\\
      \hline
      H$_2$CO & cc-pVTZ &  C$_{2v}$ & 86 & 12& 1035\\
      ScO &  VDZ (Trail-Needs ecp) & C$_{2v}$ & 76 & 17 & 830  \\
      Cr$_2$ &   cc-pVDZ-DK (x2c) &  D$_{\infty h}$ & 76 & 28 & 268 \\

    \end{tabular}
  \end{ruledtabular}
  \label{tab:molecule_info}
\end{table}

\subsection{Uncoupled optimization}
\label{uncoupled_optimization}

Uncoupled optimization separately optimizes the CI and orbital parameters in an alternating scheme:
One first selects the determinants and constructs the Hamiltonian matrix, which is diagonalized to obtain the variational wave function.
The orbital gradient ${\bf g}_o$ and possibly Hessian ${\bf h}_{oo}$ are constructed from the wave function to arrive at the new set of orbital parameters.
Finally the electronic integrals are rotated for the next iteration.

\begin{figure}[htb]
  \centerline{\includegraphics[width=0.8\textwidth]{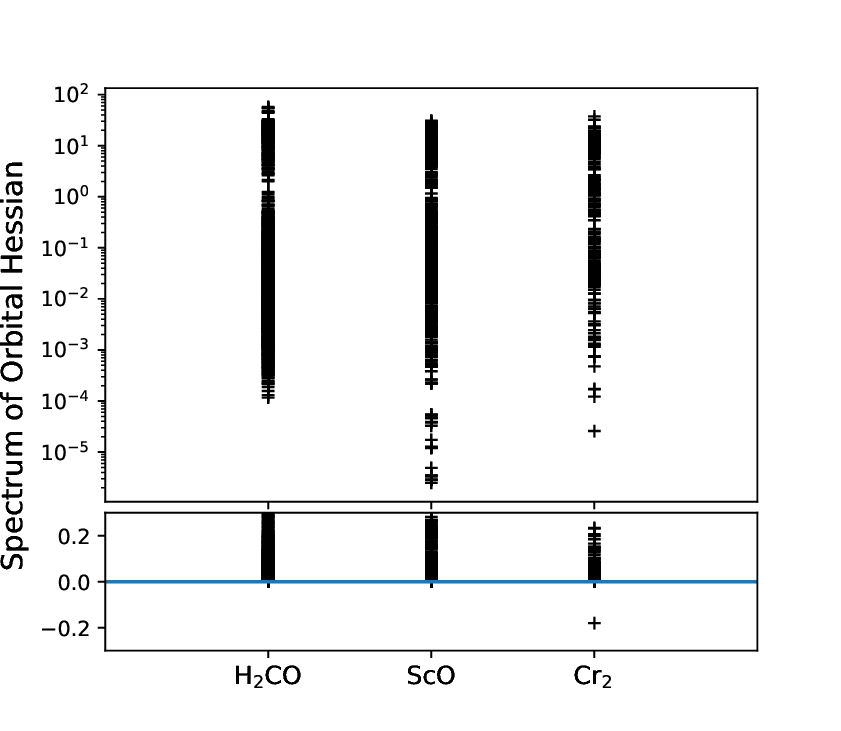}}
  \caption{The spectrum of the orbital Hessian matrix ${\bf h}_{oo}$ typically spans several orders of magnitude, with a negative
  eigenvalue in the case of Cr$_2$. The top panel shows the positive part of the spectrum on a logarithmic scale, and the bottom panel shows an expanded view of the lower part of the spectrum on a linear scale to make the negative eigenvalue visible. The Hessian matrices shown here are constructed using natural orbitals.}
  \label{fig:spectrum}
\end{figure}

Due to the large range of eigenvalues of the orbital Hessian shown in Fig.~\ref{fig:spectrum}, first-order steepest descent methods are
extremely inefficient here. Therefore, we only consider preconditioned steepest descent methods such as Newton's method.

Since we always start the optimization process from natural orbitals, the starting set of orbital parameters is already reasonably close to optimal. For this
reason, the negative eigenvalues in ${\bf h}_{oo}$ are very few and small in magnitude -- for most systems none at all.
(Note that if there
are redundant rotation parameters in the antisymmetric matrix $\bf X$, as is the case for ScO in Table~\ref{tab:molecule_info} which does not
use the full point group symmetry of the molecule, then the smallest eigenvalue of ${\bf h}_{oo}$ will be no higher than zero.)

Hence, directly solving the level-shifted Newton's equation for the orbital update $\Delta{\bf x}$ usually does not pose any problem:
\begin{align}
\label{eq:level_shifted_Newton}
({\bf h}_{oo}+\mu {\bf 1}) \Delta{\bf x} = -{\bf g}_o,
\end{align}
where $\mu\ge0$ is a small level shift chosen to overcome any negative or zero eigenvalues and ensure the positive definiteness of ${\bf h}_{oo}$. These eigenvalues can be obtained with a simple eigenvalue solver as ${\bf h}_{oo}$ is small enough. The level
shift ensures that the update is always in a descent direction in the sense that ${\bf g}_o^T \Delta{\bf x} =-{\bf g}_o^T({\bf h}_{oo}+\mu {\bf 1})^{-1}
{\bf g}_{o}<0$. As $\mu$ is increased, the parameter variations $\Delta \bf x$ become smaller and rotate from the Newton direction to the steepest
descent direction. Equivalently, $\mu$ sets a radius for the trust region to which $\Delta{\bf x}$ is restricted.

\begin{figure*}[htb]
  \centerline{\includegraphics[width=1.15\textwidth]{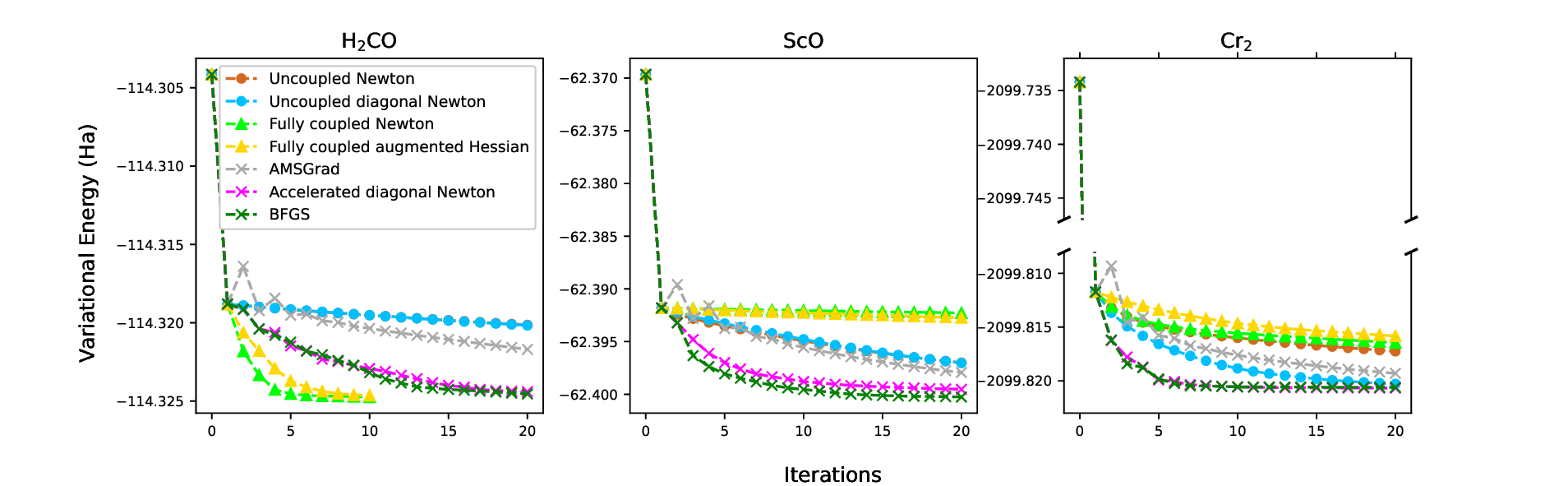}}
  \caption{Comparison of various optimization methods on the three representative systems.
  For H$_2$CO and ScO, the uncoupled Newton and uncoupled diagonal Newton curves are nearly coincident.}
  \label{fig:summary}
\end{figure*}

In quantum chemistry problems the orbital Hessian ${\bf h}_{oo}$ tends to be diagonally dominant, which is to say the coupling between the orbital
parameters is relatively weak. Therefore, instead of calculating all elements of ${\bf h}_{oo}$, we choose to just evaluate its diagonal elements.
Any eigenvalues less than $10^{-5}$ are simply set to $10^{-5}$, to ensure descent.
As shown in Fig.~\ref{fig:summary}, in practice the diagonal
approximation does not worsen the quality of the updates. In the case of Cr$_2$, the diagonal approximation performs even better than
the non-positive definite exact orbital Hessian since the level shift parameter $\mu$ inevitably slows down convergence in the descent directions.

The convergence rates of uncoupled Newton's methods can be quite slow, far from the typical
quadratic convergence behavior expected of second-order methods (see Fig.~\ref{fig:summary}). Therefore, even though we use second-order optimization in the orbital
subspace, this uncoupled scheme is not a truly second-order method due to substantial coupling between the orbital and CI parameters.
On the other hand the coupling is not so strong as to prevent convergence altogether.

\subsection{Fully coupled optimization}
\label{fully_coupled_optimization}
The slow convergence rate of uncoupled methods suggests, as has been observed in CASSCF settings, that it is important to take into account
the coupling between CI and orbital parameters in order to achieve rapid convergence. Even though these fully coupled methods can update both the CI coefficients and the orbital parameters, we instead update the CI coefficients through Hamiltonian diagonalization at the next iteration since this ensures reaching the minimum in the CI subspace.



\subsubsection{Newton's method with level shift}

As before, we can write down Newton's equation but in the entire parameter space consisting of both the $N_{\rm det}-1$ CI parameters $\bf c$
and the $N_{\rm param}$ orbital parameters $\bf x$:
\begin{align}
\label{full_newton_equation}
({\bf h}+\mu {\bf 1})
\Delta {\bf z}
= -{\bf g},
\end{align}
where
\begin{align}
{\bf h}=\begin{bmatrix}
{\bf h}_{cc} & {\bf h}_{co} \\ {\bf h}_{co}^T & {\bf h}_{oo}
\end{bmatrix},~\Delta {\bf z}=
\begin{bmatrix}
\Delta{\bf c} \\ \Delta {\bf x}
\end{bmatrix},
~{\rm and}~ {\bf g} =
\begin{bmatrix}
{\bf g}_{c} \\ {\bf g}_{o}
\end{bmatrix}.
\end{align}
This set of $N_{\rm det}+N_{\rm param}-1$ linear equations can be solved with an iterative method such as the preconditioned conjugate gradient
method. The level shift parameter $\mu\ge 0$ should be chosen to overcome all negative eigenvalues of ${\bf h}$, which ensures that the
update direction is good and that the conjugate gradient method converges.

However, even though ${\bf h}_{cc}$ is strictly positive definite as will be shown in Section \ref{full_g_h} and ${\bf h}_{oo}$ is mostly
positive definite as observed in the previous section, the full Hessian matrix ${\bf h}$ can have large negative eigenvalues due to large
magnitudes in the ${\bf h}_{co}$ block.
In such cases, the choice of the level shift parameter can be difficult without explicitly
finding the lowest eigenvalue of $\bf h$.

\subsubsection{Augmented Hessian method}

Another way of removing all negative eigenvalues in $\bf h$ is the augmented Hessian method. In this method, the dimensionality of
the Hessian matrix is enlarged by one with the gradient vector ${\bf g}$ added to the first row and column and zero added on the diagonal. The
update is determined by solving for the lowest eigenvector in the following equation:
\begin{align}
\label{ah_eqn}
\begin{bmatrix}
0 & \lambda {\bf g}^T \\
\lambda {\bf g} & {\bf h}
\end{bmatrix}
\begin{bmatrix}
1/\lambda \\
\Delta {\bf z}
\end{bmatrix} = \epsilon
\begin{bmatrix}
1/\lambda \\
\Delta {\bf z}
\end{bmatrix},
\end{align}
which multiplies out to yield
\begin{align}
\epsilon = \lambda^2 {\bf g}^T\Delta {\bf z} \\
\label{eq:eps}
({\bf h}-\epsilon {\bf 1}) \Delta {\bf z} = -{\bf g}.
\end{align}
Eq. (\ref{eq:eps}) is identical to Eq. (\ref{eq:level_shifted_Newton}) if $-\epsilon=\mu$.
Note that, since the Hessian matrix ${\bf h}$ is a principal submatrix of the augmented Hessian matrix,
according to Cauchy's interleaving theorem for Hermitian matrices,
the lowest eigenvalue $\epsilon$ of the augmented Hessian matrix
is lower than the lowest eigenvalue of ${\bf h}$ itself.
Furthermore, $\epsilon < 0$ because the first element of the augmented matrix is zero.
This guarantees that the conditioning of the Hessian matrix will be improved and all its negative eigenvalues eliminated.
Hence, the proposed move is in a downward direction, but it can overshoot.
Convergence can always be enforced with a sufficiently large $\lambda$. In addition, as convergence is approached ${\bf g}$
tends towards zero and the magnitude of $\epsilon$ also decreases.
In this respect, the augmented Hessian method is similar to the level-shifted Newton method, but with the
advantage that the level shift adjusts automatically and disappears at convergence.

Eq. (\ref{ah_eqn}) can be diagonalized with an iterative solver such as the Davidson or Lanczos method. A practical challenge is that solving this
eigenvalue problem can take many more iterations than diagonalizing the original Hamiltonian matrix -- not only is the augmented Hessian matrix
larger and denser than the original Hamiltonian matrix, it also has many eigenvalues clustered towards the lower end of its spectrum,
a property already evident in the spectrum of ${\bf h}_{oo}$ in Fig.~\ref{fig:spectrum}. In many cases, diagonalizing this matrix is painfully slow.

\subsubsection{Limitations of fully coupled methods}

The evaluation of all blocks of the Hessian matrix is prohibitively expensive. As demonstrated in detail in Section \ref{full_g_h}, the ${\bf h}_{cc}$
block has the same sparsity pattern as the Hamiltonian matrix and can be readily obtained from the Hamiltonian matrix. The ${\bf h}_{co}$
block is dense and has $(N_{\rm det}-1) \times N_{\rm param}$ elements, which can easily exceed the number of nonzero elements of the Hamiltonian matrix. Although we
present a way to construct it simultaneously with the RDMs in Section \ref{full_g_h}, it still incurs a prohibitive computational cost. As for
the ${\bf h}_{oo}$ block, while approximating it as a diagonal matrix works well in the uncoupled optimization scheme, such an approximation does not
work well for the fully coupled schemes described in this section, so one has no choice but to calculate all its elements, which is again expensive.

A workaround solution for reducing the memory usage and time complexity is to optimize only a subset of the orbital rotation parameters based
on the memory and time budget. At each iteration, one optimizes only the subset of parameters projected to give the greatest reduction in
energy. This can be achieved with a modified workflow as follows:
After the variational stage, first the RDMs are generated to construct ${\bf g}_o$
and the diagonal of ${\bf h}_{oo}$.  Then the parameter update for orbital parameter $i$ can be approximated as $-{\bf g}_{o,i}/{\bf h}_{oo,ii}$ as in the diagonal Newton's method in Section \ref{uncoupled_optimization}, and one can use a
second-order Taylor expansion to predict the resulting change in energy due to this parameter alone, namely $-{\bf g}_{o,i}^2/(2{\bf h}_{oo,ii})$.
The set of orbital parameters with the largest predicted energy contributions will be the active parameters for this iteration, so only ${\bf h}_{co}$
and ${\bf h}_{oo}$ elements corresponding to these orbital parameters need to be evaluated and stored. The rationale for this way of partitioning
the orbital parameters is that the coupling among the orbital parameters is relatively weak whereas their coupling to
CI coefficients is strong.

A more fundamental limitation comes from the fact that although the CI subspace is strictly convex and the orbital subspace is approximately
convex, the space consisting of both subspaces can be highly nonconvex.
This can result in a large number of local minima in the energy landscape.
When this occurs, the level shift added to the Hessian matrix ${\bf h}$ will
need to be very large in order to ensure convergence. Often, such very large level shifts completely overwhelm the meaningful information
from the other dimensions of the Hessian and result in very small step sizes. When this happens, convergence becomes painfully slow or leads to a local minimum.
In Fig.~\ref{fig:summary}, H$_2$CO does not have negative eigenvalues in $\bf h$ (the lowest eigenvalue being $1.79\times 10^{-6}$) and
the fully coupled Newton and augmented Hessian methods give rapid convergence.
The more difficult systems of ScO and Cr$_2$ both have a very nonconvex energy landscape (the lowest eigenvalues in $\bf h$ being -7.21 and -0.51, respectively) and much larger values of $\mu$ and $\lambda$ are needed to achieve convergence and they only converge
slowly.


\subsection{Quasi-fully coupled optimization}

Quasi-fully coupled methods take the coupling between the CI and orbital subspaces into account without computing the full Hessian $\bf h$.
These methods represent varying degrees of compromise between uncoupled and fully coupled methods.
In particular, the three methods we present in this section all require only the derivative information in the orbital subspace, whose dimensionality is usually several orders of magnitude smaller than the CI subspace, thereby incurring only a moderate computational cost.

\subsubsection{Momentum-based gradient descent methods}

AMSGrad is one of a family of momentum-based gradient descent methods commonly used in machine learning~\cite{RedKalKum-ICLR-18}. It avoids expensive
Hessian calculations since only gradient information is needed.
At each iteration $t$, for each parameter $i$ separately, AMSGrad preconditions gradient descent using running averages of the $i$-th gradient component and its square, determined
by the mixing parameters $\beta_1, \beta_2 \in (0,1)$, according to
\begin{align}
{\bf m}_i^t &= \beta_1 {\bf m}_i^{t-1}+(1-\beta_1){\bf g}_{o,i}^t, \nonumber\\
v_i^t &= \beta_2 v_i^{t-1}+(1-\beta_2){{\bf g}_{o,i}^t}^2, \nonumber\\
\hat{v}_i^t &= \max(\hat{v}_i^{t-1}, v_i^t), \nonumber\\
{\bf x}_i^{t+1} &= {\bf x}_i^{t} - \frac{\eta}{\sqrt{\hat{v}_i^t} + \epsilon} {\bf m}_i^t.
\end{align}
The parameters $\eta, \beta_1,$ and $\beta_2$ together determine the level of aggressiveness of
descent and $\epsilon$ is a small constant for numerical stability. For a variety of systems we have found that the parameters $\eta=0.01, \beta_1=0.5,\beta_2=0.5$
give reasonably good convergence, even though they are quite different from the values recommended in
the literature.

As can be seen in Fig.~\ref{fig:summary}, with these parameters
AMSGrad oscillates for the
first few iterations but eventually descends at a similar or quicker pace per iteration compared to uncoupled methods. However, the improvement is still less than satisfactory.

\subsubsection{Accelerated diagonal Newton method}

In the uncoupled optimization scheme, one observation is that adjacent orbital updates $\Delta {\bf x}$ typically point in similar directions.
This has motivated us to develop a heuristic overshooting method that achieves accelerated convergence for
most systems. Here, the overshooting tries to account for the coupling between CI and orbital parameters.

At each iteration $t$, a diagonal Newton step is calculated for the orbital parameters, but, instead of using the proposed step, it is amplified by
a factor $f^t$ determined by the cosine of the angle between the current update direction ${\bf x}^{t+1}-{\bf x}^{t}$ and
the momentum-averaged previous update direction ${\bf m}^t = \beta \frac{{\bf x}^{t} - {\bf x}^{t-1}}{\| {\bf x}^{t}-{\bf x}^{t-1}\|}  + (1-\beta) \frac{{\bf m}^{t-1}}{\| {\bf m}^{t-1}\|} $ for $\beta \approx0.5$:

\begin{align}
f^t=\min \left( \frac{2}{1-\cos({\bf m}^{t},{\bf x}^{t+1}-{\bf x}^{t})},\frac{1}{\epsilon}\right),
\end{align}
where $\epsilon$ is initialized to $0.01$ and $\epsilon\leftarrow \epsilon^{0.8}$ each time
$\cos({\bf x}^{t}-{\bf x}^{t-1},{\bf x}^{t+1}-{\bf x}^{t}) <0$.
The cosine in the expression is calculated in a ``scale-invariant" way to make it invariant under
a rescaling of some of the parameters,
i.e., in the usual definition $\cos({\bf v},{\bf w})=\langle{\bf v},{\bf w}\rangle/\sqrt{\langle{\bf v},{\bf v}\rangle \langle{\bf w},{\bf
w}\rangle}$
we define the inner product as $\langle{\bf v},{\bf w}\rangle = {\bf v}^T {\bf h}_{oo} {\bf w}$,
where the Hessian ${\bf h}_{oo}$ can again be approximated by its diagonal.
Another scale invariant choice for the inner product is
$\langle{\bf v},{\bf w}\rangle = {\bf v}^T {\bf g}_{o} {\bf g}_{o}^T {\bf w}$, and that works equally well.

This accelerated scheme exploits the approximate biconvexity of the parameter space and accounts for the coupling between the CI and orbital
subspaces. Newton steps in the orbital subspace tend to point in better directions for the orbital parameters than Newton steps in the full
parameter space, since the high degree of non-convexity in the full parameter space presents a major challenge for second-order methods. Since we re-select the determinants and reach the exact minimum in the CI subspace in each iteration, there is also greater tolerance for
overshooting the optimum in the orbital subspace since the optimal determinants and CI coefficients for that set of orbital parameters will be selected.
If the energy landscape consisting of both the CI and orbital parameters is a narrow curving valley, even though the amplified orbital parameter changes may
result in a move up the canyon wall and a high energy, on the next CI optimization step one returns
to the valley floor in the CI space and the amplified orbital move can result in faster convergence
to the minimum.
It is possible that such a scheme is more generally applicable to any situation where one alternates
between changing two sets of parameters, provided that each time one of the sets of parameters is changed
it attains the minimum in that subspace.

As can be seen in Fig.~\ref{fig:summary}, the accelerated diagonal Newton method converges
in many fewer iterations than the uncoupled methods but
not as quickly as the fully coupled methods when they work well.  However, it is the most efficient of the methods presented so far when one takes into account the per-iteration cost.


\subsubsection{The BFGS method}

In solving Eq. (\ref{full_newton_equation}), as we are only interested in the orbital updates $\Delta {\bf x}$ and not the CI updates $\Delta
{\bf c}$---the minimum in the CI subspace is reached in one step with Hamiltonian diagonalization---we could use the block matrix inversion
formula to rewrite the equation as an equation involving $\Delta {\bf x}$ only:
\begin{align}
({\bf h}_{oo}-{\bf h}_{co}^T {\bf h}_{cc}^{-1} {\bf h}_{co}) \Delta {\bf x} = -{\bf g}_o + {\bf h}_{co}^T {\bf h}_{cc}^{-1} {\bf g}_c,
\end{align}
Since ${\bf g}_c$ vanishes after the variational stage (see Section \ref{full_g_h}), the above equation simplifies to
\begin{align}
\label{bfgs_newton}
\tilde{\bf h}_{oo}  \Delta {\bf x} = -{\bf g}_o .
\end{align}
In other words, one recovers the form of Newton's equation in the orbital subspace with an effective orbital Hessian $\tilde{\bf h}_{oo} =
{\bf h}_{oo}-{\bf h}_{co}^T {\bf h}^{-1}_{cc} {\bf h}_{co}$, which contains information regarding the coupling between the CI and orbital parameters.

The Broyden-Fletcher-Goldfarb-Shanno (BFGS) method has been used for optimization in CASSCF settings.\cite{KreWerKno-JCP-19,KreWerKno-JCP-20,LevHaiTubLehWhaHea-JCTC-20} It allows one to iteratively build up approximations to the effective Hessian $\tilde{\bf h}_{oo}$ using only the orbital gradient.
For each iteration $t$, a rank-two update to $\tilde{\bf h}_{oo}$ is made using the changes in the gradient and orbital parameters:
\begin{align}
\label{bfgs_update}
\tilde{\bf h}_{oo}^{t} = \tilde{\bf h}_{oo}^{t-1}+ \frac{{\bf y}^t {{\bf y}^t}^T}{{{\bf y}^t}^T \Delta{\bf x}^{t-1}} - \frac{\tilde{\bf
h}_{oo}^{t-1} \Delta {\bf x}^{t-1}
\left(\tilde{\bf h}_{oo}^{t-1} \Delta {\bf x}^{t-1}\right)^T }{ {\Delta{\bf x}^{t-1}}^T\tilde{\bf h}_{oo}^{t-1} \Delta{\bf x}^{t-1}},
\end{align}
where
\begin{align}
{\bf y}^t={\bf g}_o^t-{\bf g}_o^{t-1}.
\end{align}

The BFGS method avoids the cost of computing and storing the numerous elements of ${\bf h}_{co}$ as well as the high cost of computing ${\bf
h}_{oo}$ by gradually building up reasonable approximations to the effective Hessian $\tilde{\bf h}_{oo}$ which contains information about
all blocks of the full Hessian matrix $\bf h$.
Furthermore, the BFGS update rule (\ref{bfgs_update}) preserves the positive definiteness of $\tilde{\bf h}_{oo}$ as long as ${{\bf y}^t}^T \Delta{\bf x}^{t-1}>0$, which is guaranteed because
we skip those updates for which either of the denominators in Eq. (\ref{bfgs_update}) are small
(say below $10^{-5}$).
For the initial Hessian $\tilde{\bf h}^0_{oo}$, we can use the diagonal of ${\bf h}_{oo}^0$ as an approximation with
any negative elements reset as usual.
When solving the linear system in Eq. (\ref{bfgs_newton}) we add also a small diagonal constant
(typically $10^{-3}$) to $\tilde{\bf h}_{oo}$ to ensure that its spectrum is sufficiently positive.

As shown in Fig. \ref{fig:summary}, the BFGS method achieves similar convergence rates as accelerated diagonal Newton with a lower cost per iteration, since, after the first iteration, only the gradient needs to be evaluated.

\subsection{Microiterations}

\begin{figure*}[htb]
	\centerline{\includegraphics[width=1.15\textwidth]{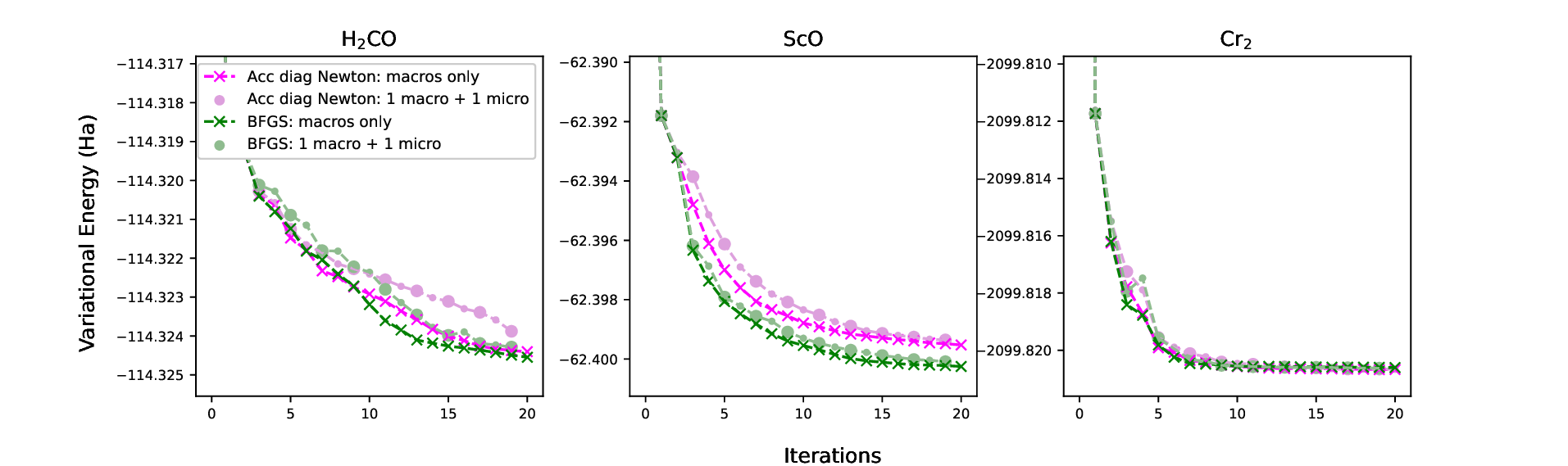}}
	\caption{The comparison of alternating macro- (large dots) and micro-iterations (small dots) with a macroiterations-only scheme. }
	\label{fig:micro}
\end{figure*}

The advantage of re-selecting the determinants for each iteration as we have done thus far is that we always work with the optimal set of
determinants for a given set of orbital parameters. However, the process of selecting the determinants and constructing the corresponding
Hamiltonian matrix can be a costly procedure in itself, especially for systems with a large number of electrons. (In comparison, the construction
of orbital gradient and Hessian as well as the rotation of integrals is independent of $N_{\rm elec}$ and only scales with $N_{\rm orb}$.) Here,
we investigate the option of having microiterations in which the determinants from the previous iteration are kept.
This way, instead of going through the expensive step of constructing a new Hamiltonian matrix,
we only need to go through all the nonzero elements of the sparse Hamiltonian matrix and update them using the
new set of integrals, incurring a cost similar to constructing the two-body reduced density matrix (2-RDM) detailed in Section \ref{RDMs}.
Then the Davidson method is used to find the new lowest eigenvector(s), using the saved eigenvector(s) from
the previous iteration as a starting point.
The determinants are only re-selected during the next macroiteration.

As shown in Fig.~\ref{fig:micro}, including one microiteration after each macroiteration does slow down convergence slightly compared to the
macroiterations-only scheme presented previously, but the saving in the amount of work per iteration more than makes up for the difference, especially for the systems with a larger number of electrons.

With microiterations we can also get a sense of how much the reselection of determinants contributes to the gain in energy.
If we disable the reselection of determinants after the natural orbital step, then after 20 BFGS microiterations
we eventually recover only 24\%, 47\%, and 63\% of the total energy gain for the three systems respectively.
Moreover, as seen from Fig.~\ref{fig:extrapolation} when macroiterations are employed there is frequently
(though not always) a reduction in the number of determinants selected for a given $\epsilon_1$, which of course is absent
in the microiterations-only scheme.

\subsection{State-averaged optimization}

\begin{figure*}[htb]
	\centerline{\includegraphics[width=1.15\textwidth]{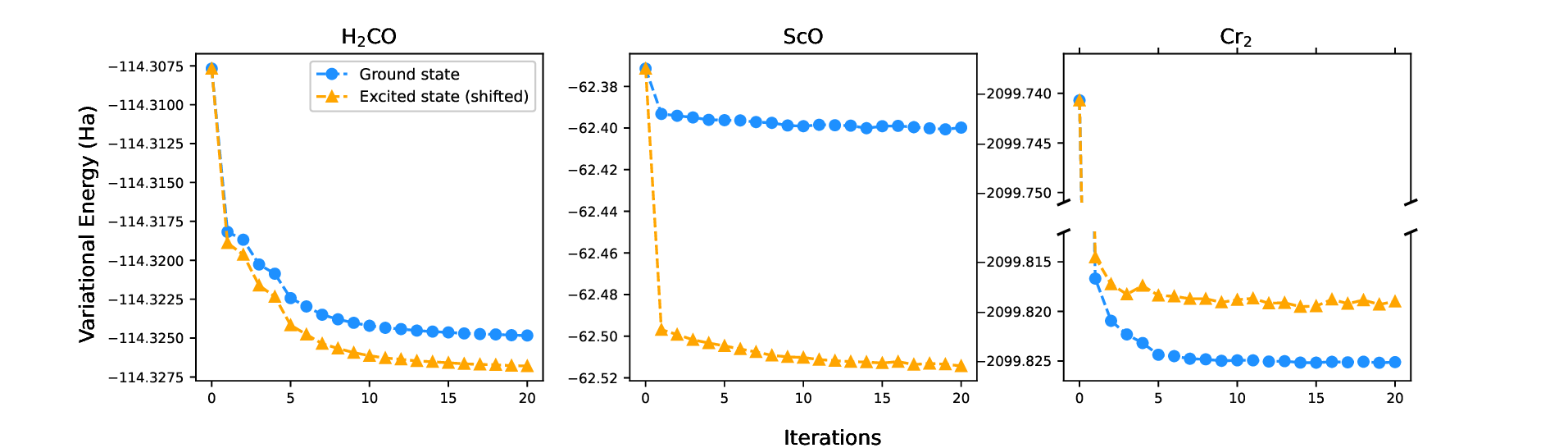}}
	\caption{State-averaged optimization of the ground and first excited states with equal weighting for the two states using the accelerated diagonal Newton method. For ease of comparison, the excited state is shifted downward so that it has the same starting energy as the ground state.
}
	\label{fig:excited}
\end{figure*}

When excited state wave functions are desired, optimizing only the excited state energy could become unstable. In this case, one can optimize
the (weighted) sum of the energies of the ground and excited states. This is called state-averaged optimization, illustrated in Fig. \ref{fig:excited}.

The computational time of state-averaged optimization does not increase linearly with the number of states being optimized when, as is usually the case, only a small number of excited states are computed.  After the
variational stage, where the Davidson diagonalization solver now needs to converge not only the ground state  but also as many excited
states as are needed, one can simply evaluate the RDMs as the sum of the RDMs for each state as the variational energy is a linear function of
the RDMs. Therefore, the computationally expensive construction of the gradient and Hessian does not change at all,
but the number of determinants needed to reach a given ground state variational energy increases, as does
the time for Hamiltonian construction and diagonalization.

\subsection{Comparison of Natural Orbitals and Optimized Orbitals}

\begin{figure*}[htb]
	\centerline{\includegraphics[width=1.05\textwidth]{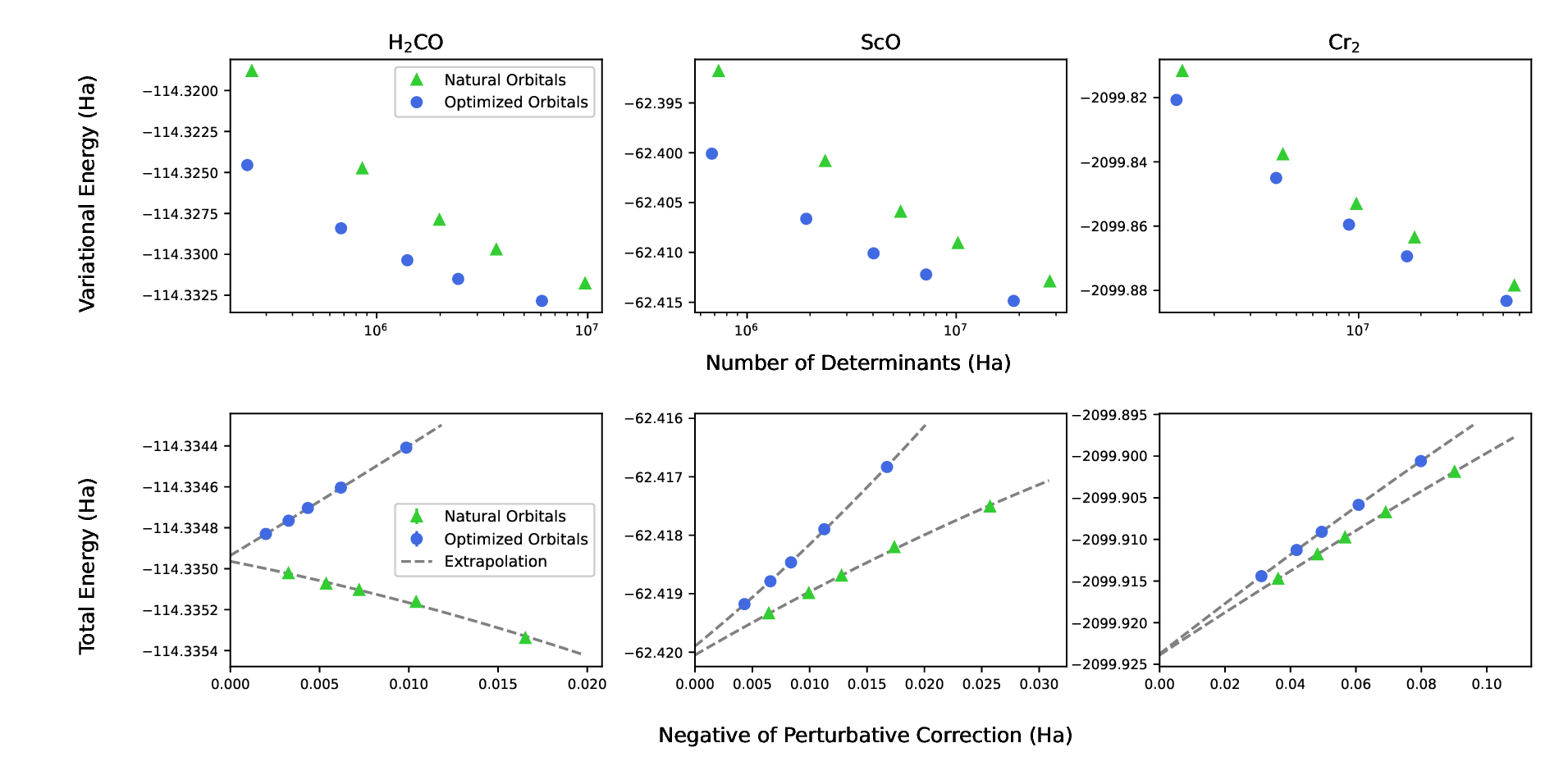}}
	\caption{Comparison of the performance of natural orbitals and optimized orbitals. The top row displays the variational energy $E_V$ versus the number of determinants in the CI expansion $N_{\rm det}$. The bottom row displays the extrapolation of SHCI total energies to the FCI limit, i.e.   $\epsilon_1 \rightarrow0$ and $\Delta E^{(2)} \rightarrow0$.
The natural orbitals are obtained from the initial SHCI wave function
and the optimized orbitals are obtained from BFGS optimization, both at $\epsilon_{1}=2 \times 10^{-4}$.
The calculations are performed at four more $\epsilon_{1}$ values: $1 \times 10^{-4}$, $6 \times 10^{-5}$, $4 \times 10^{-5}$, and $2 \times 10^{-5}$. The perturbative calculations use $\epsilon_{2}=\epsilon_{1}\times 10^{-3}$. Optimized orbitals are clearly superior to natural orbitals in producing more compact variational wave functions and reducing the magnitude of the perturbative correction.}
	\label{fig:extrapolation}
\end{figure*}

In Fig. \ref{fig:extrapolation}, we compare the performance of natural orbitals and optimized orbitals for the three systems.
These  orbitals are optimized at an intermediate value of $\epsilon_{1}$, in this case $2 \times 10^{-4}$, using 18 iterations of BFGS optimization after natural orbitals, as in Fig. {\ref{fig:summary}}.
The final calculations start from this value of $\epsilon_{1}$ and go down to four more values: $1 \times 10^{-4}$, $6 \times 10^{-5}$, $4 \times 10^{-5}$, and $2 \times 10^{-5}$. The perturbative calculations use $\epsilon_{2}= \epsilon_{1}\times 10^{-3}$.

As shown in the top row of Fig. \ref{fig:extrapolation}, the optimized orbitals produce much more compact wave functions for the same variational energy. As a major memory bottleneck with SCI methods is the storage of the sparse Hamiltonian matrix whose size scales super-linearly and sub-quadratically with the number of determinants, orbital optimization can significantly alleviate the memory bottleneck. The optimized orbitals show the greatest advantage over natural orbitals at $\epsilon_{1}=2 \times 10^{-4}$, but this advantage also carries over to smaller $\epsilon_{1}$ values, although to a lesser degree. The bottom row shows that this improvement in orbital quality also reduces the magnitude of SHCI perturbative correction.

Of course, the smallest $\epsilon_1$ value matters the most for reducing extrapolation errors. To see even greater benefit of orbital optimization, one could choose to optimize the orbitals at an $\epsilon_{1}$ value closer to the smallest one, at a higher per-iteration computational cost.
However, we find for these systems that optimizing the orbitals using a value of $\epsilon_1$ smaller than $2 \times 10^{-4}$
does not result in much further improvement.

\section{Evaluation of various pieces}
\label{evaluation_pieces}

In this section, we explain in detail how to efficiently evaluate various pieces used in the previous section.
To simplify notation, we omit the subscript and use $|\Psi\rangle $ to denote the variational wave function. As in the rest of this paper, we assume real wave functions, i.e., wave functions with real CI coefficients but not necessarily real orbitals.

\subsection{Reduced density matrices}
\label{RDMs}
The 1-RDM is used both in constructing the natural orbitals as a starting point for further optimization, and in evaluating the orbital gradient
and Hessian elements.

The 1-RDM is defined as
\begin{align}
D_{pq}=\sum_{\sigma} \langle \Psi | a_{p\sigma}^{\dagger} a_{q\sigma} | \Psi \rangle.
\end{align}

A fast way of constructing the 1-RDM is as follows: First loop over all determinants in the variational wave function and generate all
possible single excitations on them; if a resulting determinant is also in the wave function,  add to the appropriate entry of the 1-RDM. This
approach has overall time complexity $O(N_{\rm det} N_{\rm elec} N_{\rm  orb})$.

The 2-RDM is defined as
\begin{align}
\label{2rdm_defn}
d_{pqrs}=\sum_{\sigma\tau} \langle \Psi | a_{p\sigma}^{\dagger} a_{r\tau}^{\dagger} a_{s\tau} a_{q\sigma} | \Psi \rangle.
\end{align}

We studied two approaches for constructing the 2-RDM. The first approach is similar to the one we used to construct the 1-RDM, i.e., loop over
all determinants and generate all possible excitations.
This approach has time complexity $O(N_{\rm det} N_{\rm  elec}^2 N_{\rm  orb}^2)$ and can be quite slow.

A more efficient approach makes use of the Hamiltonian matrix from the variational stage and the fact that for the quantum chemistry Hamiltonian in Eq. (\ref{qc_hamiltonian})
determinant pairs that contribute to the 2-RDM coincide with pairs that have nonzero matrix elements in the Hamiltonian matrix. In other words,
one only needs to iterate over all nonzero elements in the Hamiltonian matrix and do the following:
\begin{enumerate}
\item For each identical pair of determinants, loop over all $O(N_{\rm elec}^2)$ ways they can be connected by the four fermion operators in Eq. (\ref{2rdm_defn}) and add the contributions to the relevant entries of the 2-RDM.

\item For each singly excited pair, loop over all $O(N_{\rm elec})$ ways they can be connected and update the 2-RDM.

\item For each doubly excited pair, loop over the $O(1)$ ways they can be connected and  update the 2-RDM.
\end{enumerate}

Symmetry of the 2-RDM,
\begin{align}
d_{pqrs}=d_{rspq}=d_{qpsr}=d_{srqp},
\end{align}
can be used to reduce storage.

Additionally, when both 1- and 2-RDMs are needed, one can first evaluate the 2-RDM and then obtain the 1-RDM more quickly with $O(N_{\rm orb})$ cost for each element through index contraction:
\begin{align}
D_{pq}=\frac{1}{N_{\rm elec}-1}\sum_{m} d_{pqmm}.
\end{align}
Hence, the total cost of constructing the 1-RDM reduces to $O(N_{\rm orb}^3)$.

\subsection{Orbital gradient and Hessian}

The construction of the orbital gradient ${\bf g}_o$ and  Hessian ${\bf h}_{oo}$ are reviewed in Ref. \onlinecite{HelJorOls-BOOK-02}. They can be  obtained through a Taylor expansion of the variational energy:
\begin{align}
\label{energy}
E({\bf X}) &= \langle \Psi | \exp(\hat{{\bf X}}) \hat{H} \exp(-\hat{{\bf X}}) | \Psi \rangle \nonumber \\
&= E_{V} + {\bf x}^T {\bf g}_{o} + \frac{1}{2} {\bf x}^T {\bf h}_{oo}{\bf x} + \cdots.
\end{align}
For notational clarity, the $N_{\rm param} \times 1$ vectorized form of the upper triangular part of the antisymmetric matrix $\bf X$ has been
denoted $\bf x$. The elements of $\bf x$ are indexed with composite orbital indices such as $pq$ where $p$ and $q$ are the original row and
column indices in the matrix $\bf X$.

We can then identify the corresponding gradient and Hessian elements at ${\bf x} = {\bf 0}$:
\begin{align}
{\bf g}_{o,pq}&= \frac{\partial E({\bf 0})}{\partial {\bf x}_{pq}} = \big\langle\Psi\big|\big[\hat{E}_{pq}^{-}, \hat{H}\big] \big| \Psi\big\rangle \label{go_1}\\
{\bf h}_{oo,pqrs} &= \frac{\partial^2 E({\bf 0})}{\partial {\bf x}_{pq} \partial {\bf x}_{rs}} = \frac{1}{2}(1+ P_{pq,rs})
\big\langle \Psi \big| \big[\hat{E}_{pq}^{-}, \big[\hat{E}_{rs}^{-},\hat{H}\big]\big] \big| \Psi \big\rangle. \label{Hoo_1}
\end{align}
$\hat{E}_{pq}^{-}$ has been defined in Eq. (\ref{orbital_excitation}).
$P_{pq,rs}$ permutes the indices $pq$ and $rs$, thereby symmetrizing the Hessian.

The orbital gradient in Eq. (\ref{go_1}) can be written as
\begin{align}
{\bf g}_{o,pq} = 2(F_{pq}-F_{qp})
\end{align}
where the generalized Fock matrix is
\begin{align}
F_{mn} &= \sum_{\sigma} \langle\Psi| a_{m\sigma}^{\dagger}\left[a_{n\sigma},\hat{H}\right] |\Psi\rangle.
\end{align}
After some algebra, one can rewrite $F$ as the contraction of the RDMs with the electronic integrals.
\begin{align}
\label{generalized_Fock2}
F_{mn} = \sum_{q}D_{mq}h_{nq}+\sum_{qrs}d_{mqrs}g_{nqrs}.
\end{align}
Therefore, calculating each element of the orbital gradient takes $O(N_{\rm orb}^3)$ work when the RDMs are available, and so construction of
the entire orbital gradient takes $O(N_{\rm orb}^5)$ work.

Similarly, the Hessian in Eq.~(\ref{Hoo_1}) can be written in terms of the RDMs and integrals as follows:
\begin{align}
{\bf h}_{oo,pqrs} &= (1-P_{pq})(1-P_{rs})A_{pqrs} \nonumber \\
&=  A_{pqrs} - A_{pqsr} - A_{qprs} + A_{qpsr},
\end{align}
where
\begin{align}
A_{pqrs} &= 2D_{pr}h_{qs} - (F_{pr} +F_{rp})\delta_{qs} +2Y_{pqrs} \\
Y_{pqrs} &= \sum_{mn} \left[(d_{pmrn}+d_{pmnr})g_{qmns} + d_{prmn} g_{qsmn}\right] \\
Y_{pqrs} &= Y_{rspq}.
\end{align}
After construction of the generalized Fock matrix, each element of the $Y$ matrix can be evaluated with $O(N_{\rm orb}^2)$ cost,
resulting in $O(N_{\rm orb}^4)$ cost for the diagonal of the Hessian, or $O(N_{\rm orb}^6)$ cost for the entire Hessian.

\subsection{Full gradient and Hessian}
\label{full_g_h}
In order to perform fully coupled second-order optimization, in addition to the orbital gradient and Hessian ${\bf g}_o$ and ${\bf h}_{oo}$, we also need the CI gradient ${\bf g}_c$,
and the ${\bf h}_{cc}$ and ${\bf h}_{co}$ blocks of the Hessian.
These derivatives should be applied to the full energy expression, Eq.~(\ref{energy2}).

The CI derivative evaluated at the current set of CI coefficients and orbital parameters yields the CI gradient:
\begin{equation}
\label{gc}
{\bf g}_{c,i} = \frac{\partial E({\bf c}, {\bf 0})}{\partial c_i} =2\langle D_i|\hat{H}|\Psi\rangle - 2c_i E_V.
\end{equation}
The CI gradient vanishes when the wave function $|\Psi\rangle$ is an eigenvector of the Hamiltonian matrix, which is the case after diagonalization
of the Hamiltonian matrix in the variational stage, $\hat{H}|\Psi\rangle = E_V |\Psi \rangle$.

Taking second derivatives of the variational energy (\ref{energy2}) gives the ${\bf h}_{cc}$ and ${\bf h}_{co}$ blocks of the Hessian:
\begin{align}
{\bf h}_{cc,ij}
&=\frac{\partial^2 E({\bf c},{\bf 0})}{\partial c_i \partial c_j} \nonumber \\
&=-4c_j\langle D_i|\hat{H}|\Psi\rangle -4c_i\langle D_j|\hat{H}|\Psi\rangle +2\langle D_i|\hat{H}| D_j\rangle \nonumber \\
&~~~~ +(8c_ic_j-2\delta_{ij}) E_{V} \nonumber \\
&=2\langle D_i|\hat{H}| D_j\rangle - 2\delta_{ij}E_{V}.
\end{align}
Therefore, ${\bf h}_{cc}$ is simply the $(N_{\rm det}-1) \times (N_{\rm det}-1)$ submatrix of the Hamiltonian matrix rescaled and shifted on
the diagonal. Since the lowest eigenvalue of the $(N_{\rm det}-1) \times (N_{\rm det}-1)$ submatrix will always be above $E_V$, ${\bf h}_{cc}$
is strictly positive definite.

\begin{align}
{\bf h}_{co,(i,pq)}
&=\frac{\partial^2 E({\bf c},{\bf 0})}{\partial c_i \partial {\bf x}_{pq}} \nonumber \\
&= 2 \langle D_i|\left[\hat{E}_{pq}, \hat{H}\right] | \Psi\rangle +2 \langle\Psi|\left[\hat{E}_{pq}, \hat{H}\right] | D_i\rangle \nonumber \\
& ~~~~ - 2c_i \langle \Psi| \left[\hat{E}_{pq}^{-},\hat{H}\right]|\Psi\rangle \nonumber \\
&= 2 \langle D_i|\left[\hat{E}_{pq}, \hat{H}\right] | \Psi\rangle -2 \langle D_i|\left[\hat{E}_{qp}, \hat{H}\right] | \Psi\rangle \nonumber \\
& ~~~~-2c_i
{\bf g}_{o,pq} \label{h_co}.
\end{align}
Unlike ${\bf h}_{cc}$, ${\bf h}_{co}$ is dense. It is also typically significantly larger in size than ${\bf h}_{oo}$. We provide an approach below for
its construction.

The third term of Eq.~(\ref{h_co}) is simply the orbital gradient. For the first term, and similarly for the second term, consider the one- and two-body parts separately:
\begin{align}
&\langle D_i|\left[\hat{E}_{pq}, \hat{H}\right] | \Psi\rangle \nonumber \\
=& \langle D_i|\left[\hat{E}_{pq}, \hat{h}\right] | \Psi\rangle+\langle D_i|\left[\hat{E}_{pq}, \hat{g}\right] | \Psi\rangle,
\end{align}
where $\hat{h}$ and $\hat{g}$ are defined in Eq. (\ref{qc_hamiltonian}).

First look at the one-body part:
\begin{align}
&\langle D_i|\left[\hat{E}_{pq}, \hat{h}\right] | \Psi\rangle \nonumber \\
=&\sum_{\sigma} \langle D_i | a_{p\sigma}^{\dagger}\left[a_{q\sigma},\hat{h}\right] | \Psi\rangle - \sum_{\sigma} \langle \Psi |
a_{q\sigma}^{\dagger}\left[a_{p\sigma},\hat{h}\right] | D_i\rangle \nonumber \\
=& \sum_{t}\sum_{\sigma} h_{qt} \langle D_i | a_{p\sigma}^{\dagger} a_{t\sigma} |\Psi\rangle - \sum_{t}\sum_{\sigma} h_{pt} \langle\Psi |
a_{q\sigma}^{\dagger} a_{t\sigma} |D_i\rangle \nonumber \\
=& \sum_{j}\sum_{t}\sum_{\sigma} c_j h_{qt} \langle D_i | a_{p\sigma}^{\dagger} a_{t\sigma} |D_j\rangle \nonumber \\
&~~~ - \sum_{j}\sum_{t}\sum_{\sigma} c_j h_{pt} \langle D_j | a_{q\sigma}^{\dagger} a_{t\sigma} |D_i\rangle.
\end{align}
And similarly the two-body part:
\begin{align}
&\langle D_i|\left[\hat{E}_{pq}, \hat{g}\right] | \Psi\rangle \nonumber \\
=&\sum_{\sigma} \langle D_i | a_{p\sigma}^{\dagger}\left[a_{q\sigma},\hat{g}\right] | \Psi\rangle - \sum_{\sigma} \langle \Psi |
a_{q\sigma}^{\dagger}\left[a_{p\sigma},\hat{g}\right] | D_i\rangle \nonumber \\
=&\sum_{tuv} \sum_{\sigma\tau} g_{qtuv} \langle D_i | a_{p\sigma}^{\dagger}a_{u\tau}^{\dagger} a_{v\tau}a_{t\sigma}|\Psi\rangle \nonumber \\
& ~~~~ -
\sum_{tuv} \sum_{\sigma\tau} g_{ptuv} \langle\Psi | a_{q\sigma}^{\dagger}a_{u\tau}^{\dagger} a_{v\tau}a_{t\sigma}|D_i\rangle \nonumber \\
=&\sum_{j}\sum_{tuv} \sum_{\sigma\tau} c_j g_{qtuv} \langle D_i | a_{p\sigma}^{\dagger}a_{u\tau}^{\dagger} a_{v\tau}a_{t\sigma}|D_j\rangle
\nonumber \\
&~~~~ - \sum_{j}\sum_{tuv} \sum_{\sigma\tau} c_j g_{ptuv} \langle D_j | a_{q\sigma}^{\dagger}a_{u\tau}^{\dagger} a_{v\tau}a_{t\sigma}|D_i\rangle.
\end{align}

The summands in the above expressions have in fact already been evaluated when calculating the RDMs. Recall that when the 1-RDM is calculated,
each nonzero permutation factor $\langle D_i | a_{p\sigma}^{\dagger} a_{q\sigma} |D_j\rangle$ gets evaluated and $c_ic_j\langle D_i | a_{p\sigma}^{\dagger}
a_{q\sigma} |D_j\rangle$ added to the $(p,q)$-entry of the 1-RDM.
When the 2-RDM is calculated, each nonzero permutation factor $\langle D_i | a_{p\sigma}^{\dagger}a_{r\tau}^{\dagger} a_{s\tau}a_{q\sigma}|D_j\rangle$
gets evaluated and $c_ic_j\langle D_i | a_{p\sigma}^{\dagger}a_{r\tau}^{\dagger} a_{s\tau}a_{q\sigma}|D_j\rangle$
added to the $(p,q,r,s)$-entry of the 2-RDM. This means that whenever a permutation factor  is evaluated during RDM construction,
we should update the corresponding elements of ${\bf h}_{co}$ accordingly, of which there are $O(N_{\rm orb})$ in number. In more detail, the one- and two-body parts of ${\bf h}_{co}$ should be updated according to Algorithms~\ref{Hco_1b} and \ref{Hco_2b} respectively, again assuming that the set of nonredundant orbital parameters lie in the upper triangular part of $\bf X$. The cost of constructing ${\bf h}_{\rm co}$ is thus a factor of $O(N_{\rm orb})$ more
expensive than the cost of constructing the RDMs.

\begin{algorithm}
  \caption{Given $\langle D_k | a_{u\sigma}^{\dagger} a_{v\sigma} |D_l \rangle$, update ${\bf h}_{co}$ according to the following rule.
}
  \label{Hco_1b}
  \begin{algorithmic}
    \FOR{$s=u+1,u+2,...,N_{orb}$}
    \IF{$u$ and $s$ belong to the same irrep}
    \STATE ${\bf h}_{co(k,us)}\mathrel{+}= 2c_l h_{sv} \langle D_k | a_{u\sigma}^{\dagger} a_{v\sigma} |D_l\rangle$
        \STATE ${\bf h}_{co(l,us)}\mathrel{+}= 2c_k h_{sv} \langle D_k | a_{u\sigma}^{\dagger} a_{v\sigma} |D_l\rangle$
    \ENDIF
    \ENDFOR
    \FOR{$s=1,2,...,u-1$}
    \IF{$u$ and $s$ belong to the same irrep}
    \STATE ${\bf h}_{co(k,su)}\mathrel{-}= 2c_l h_{sv} \langle D_k | a_{u\sigma}^{\dagger} a_{v\sigma} |D_l\rangle$
        \STATE ${\bf h}_{co(l,su)}\mathrel{-}= 2c_k h_{sv} \langle D_k | a_{u\sigma}^{\dagger} a_{v\sigma} |D_l \rangle$
    \ENDIF
    \ENDFOR

  \end{algorithmic}
\end{algorithm}

\begin{algorithm}
	\caption{Given $\langle D_k | a_{t\sigma}^{\dagger}a_{u\tau}^{\dagger} a_{v\tau}a_{w\sigma}|D_l \rangle$, update ${\bf h}_{co}$ according to the following rule.
	}
	\label{Hco_2b}
	\begin{algorithmic}

	\FOR{$s=t+1,t+2,...,N_{orb}$}
	\IF{$t$ and $s$ belong to the same irrep}
	\STATE ${\bf h}_{co(k,ts)}\mathrel{+}= 2c_l g_{swuv} \langle D_k | a_{t\sigma}^{\dagger}a_{u\tau}^{\dagger} a_{v\tau}a_{w\sigma}|D_l\rangle$
		\STATE ${\bf h}_{co(l,ts)}\mathrel{+}= 2c_k g_{swuv} \langle D_k | a_{t\sigma}^{\dagger}a_{u\tau}^{\dagger} a_{v\tau}a_{w\sigma}|D_l\rangle$
	\ENDIF
	\ENDFOR
	\FOR{$s=1,2,...,t-1$}
	\IF{$t$ and $s$ belong to the same irrep}
	\STATE ${\bf h}_{co(k,st)}\mathrel{-}= 2c_l g_{swuv} \langle D_k | a_{t\sigma}^{\dagger}a_{u\tau}^{\dagger} a_{v\tau}a_{w\sigma}|D_l\rangle$
		\STATE ${\bf h}_{co(l,st)}\mathrel{-}= 2c_k g_{swuv} \langle D_k | a_{t\sigma}^{\dagger}a_{u\tau}^{\dagger} a_{v\tau}a_{w\sigma}|D_l\rangle$
	\ENDIF
	\ENDFOR

	\end{algorithmic}
\end{algorithm}

\subsection{Orbital rotation matrix}

Once we have the parameter update $\bf x$ filled into the antisymmetric matrix $\bf X$, we would like to construct the orthogonal rotation
matrix $\exp(-\bf X)$.

One way to evaluate this exponential is through its Taylor expansion $\exp(-{\bf X}) = \sum\limits_{n=0}^{\infty} (-{\bf X})^n /n!$, truncating the
expansion at some order. This can result in loss of unitarity if the norm of ${\bf X}$ is not small.

An alternative, discussed for example in Ref.~\onlinecite{HelJorOls-BOOK-02}, is to  write ${-\bf X}$ in terms of its eigenvalues and eigenvectors:
\begin{align}
{-\bf X} = i{\bf V T V}^{\dagger}, ~~~~~ {\bf V}^{\dagger}{\bf V} = {\bf 1},
\end{align}
where $\bf T$ is a real diagonal matrix and ${\bf V}$ is complex unitary.
Then, $\exp(-{\bf X}) =  {\bf V}\exp(i{\bf T} ){\bf V}^{\dagger}$.

To avoid complex arithmetic, we can instead diagonalize the square of $-{\bf X}$:
\begin{align}
{\bf X}^2 =-{\bf W}{\bf T}^2{\bf W}^T, ~~~~~ {\bf W}^T{\bf W} = {\bf 1}.
\end{align}
Since, ${\bf X}^2$ is real and symmetric, its eigenvalues are real (and nonpositive since the eigenvalues of $-{\bf X}$ are purely imaginary or zero). Now,
\begin{align}
\exp(-{\bf X}) &=\sum_{n=0}^{\infty} \frac{1}{(2n)!}{\bf X}^{2n} + \sum_{n=0}^{\infty} \frac{1}{(2n+1)!}{\bf X}^{2n}(-{\bf X}) \nonumber \\
& = {\bf W} \cos({\bf T}) {\bf W}^T + {\bf W}{\bf T}^{-1}\sin({\bf T}) {\bf W}^T (-{\bf X}).
\end{align}

After the rotation matrix is constructed, rotating the electronic integrals can be done in $O(N_{\rm orb}^5)$ time instead of $O(N_{\rm orb}^8)$
by sequentially transforming the four indices rather than with nested loops.

\section{Conclusions}
We have studied orbital optimization in SCI methods with a particular focus on SHCI. We have presented a number of optimization schemes and demonstrated the importance of taking into account the coupling between CI and orbital parameters when designing fast converging methods. Using three representative systems we have shown that two quasi-fully coupled methods---accelerated diagonal Newton and BFGS---are the methods of choice in such applications.
Compared to natural orbitals, optimized orbitals can yield more compact representations of the variational wave function and reduce the magnitude of the perturbative correction.
We have also provided ample detail on the efficient evaluation of various quantities used in the optimization procedure.

In this paper we have limited our discussion of orbital optimization to the context of using SCI+PT methods as an approximation
to FCI.
The ideas presented here can also be useful when SCI is used for other purposes, e.g., as an approximate active space solver in CASSCF,
or to provide trial wave functions for quantum Monte Carlo.

\label{conclusions}

\begin{acknowledgements}
We thank Hans-Joachim Werner for valuable comments on the manuscript and Sandeep Sharma, Garnet Chan, and Qiming Sun for helpful discussions.
Y.Y. thanks Benjamin Pritchard for several suggestions on efficient implementation and acknowledges fellowship support from the Molecular Sciences Software Institute funded by U.S. National Science Foundation grant ACI-1547580.
This work was supported in part by the AFOSR under grant FA9550-18-1-0095.
Some of the computations were performed at the Bridges cluster at the Pittsburgh Supercomputing Center supported by NSF grant ACI-1445606.
\end{acknowledgements}

\bibliographystyle{achemso}
\bibliography{all}

\end{document}